\documentclass[12pt]{article}
\usepackage{amsmath}
\usepackage{amssymb}
\usepackage{amsfonts}

\input{tcilatex}

\begin{document}

\begin{center}
\bigskip

\textbf{Remarks on antichains in the causality order of space-time}
\end{center}

\bigskip

\begin{center}
Stephan Foldes

2017

\bigskip

\textbf{Abstract}
\end{center}

\textit{The two closely related Lorentz-invariant partial orders of
space-time are distinguished with respect to the existence of antichain
cutsets and the possibility of grading. World lines of particles with or
without mass are the maximal chains in the causality order of space-time,
and antichain cutsets are the levels of the various gradings of the
causality partial order. The maximal chains of the weaker, subluminal
causality order need not be connected topologically, subluminal causality
has no antichain cutsets and cannot be graded. Combinatorial
characterizations of optical lines and hyperplanes, separation lines,
inertia planes and lines, ultimately in terms of the causality order yield a
simple proof of the Alexandrov-Zeeman Theorem.}

\textit{\bigskip }


\bigskip

\bigskip

\textbf{1 Poset grading, levels, antichain cutsets}

\bigskip

In a graded (ranked) partially ordered set, each level (set of all elements
of the same rank) is a maximal antichain that intersects every maximal
chain. If the partially ordered set satisfies certain conditions, then the
converse is also true: in such posets, every antichain that intersects evey
maximal chain is a level under a grading of the poset (see Rival and Zaguia
[RZ] and [FW]). Grading is a natural idea in discrete posets, but also in
general, a surjective map $g$ from a poset $P$ onto a totally ordered set
(chain) $R$ may be called a \textit{grading} if its restriction to each
maximal chain $C$ of $P$ is an isomorphism from $C$\ to $R.$ A \textit{level}
with respect to this grading is then defined as the pre-image under $g$ of
any element of $R$. For example, on the distributive lattice $%
\mathbb{R}
^{n}$, the sum of vector components defines a grading $%
\mathbb{R}
^{n}\longrightarrow 
\mathbb{R}
$, which is not at all unique, but its restriction to the integer lattice
yields the essentially unique grading $\mathbf{Z}^{n}\longrightarrow \mathbf{%
Z}.$

\bigskip

In a partially ordered set, an antichain that intersects every maximal chain
is called an\textit{\ antichain cutset.} Every level of a graded poset is an
antichain cutset, and every antichain cutset is a maximal antichain.
Converse statements do not hold generally, and even in such well-behaved
posets as a Boolean lattice, maximal antichains need not be cutsets.
However, extending a result contained in [RZ], it was shown in [FW] that in
every discrete, strongly connected poset, every antichain cutset is a level
under an essentially unique grading.

\bigskip

\bigskip

\textbf{2 \ Causality order and antichains as space-like hypersurfaces }

\bigskip

For each non-negative integer $n$ and positive real number $c$, the \textit{%
causality} order $\leq _{c}$ on $(n+1)$-dimensional space-time $%
\mathbb{R}
^{n}\times 
\mathbb{R}
=%
\mathbb{R}
^{n+1}$ is given by 
\begin{equation*}
\mathbf{x}t<_{c}\mathbf{x}^{\prime }t^{\prime }\Leftrightarrow \text{ }%
t<t^{\prime }\text{ and }\frac{\left\Vert \mathbf{x}^{\prime }-\mathbf{x}%
\right\Vert }{t^{\prime }-t}\leq c
\end{equation*}%
Mathematical interest in the causality order is due in significant part to
the Alexandrov-Zeeman Theorem (see [A1], [AO], [A-CJM], [Z] and Section 4
below), which states that in at least $(2+1)$-dimensional space-time, the
automorphism group of the causality order is the semi-direct product of the
Poincar\'{e} group and the group of space-time dilations (or equivalently,
it is generated by Lorentz boosts, space rotations, and translations and
dilations of space-time). The result does not hold in $(1+1)$-dimensional
space-time, due to the paucity of space rotations. In $1+1$ dimensions the
causality poset is a distributive lattice isomorphic to the componentwise
lattice order on $%
\mathbb{R}
^{2}$, but it is not a lattice in higher dimensions.

By a \textit{world line} in $(n+1)$-dimensional space-time $%
\mathbb{R}
^{n+1}=%
\mathbb{R}
^{n}\times 
\mathbb{R}
$ we mean the inverse of the graph in $%
\mathbb{R}
\times 
\mathbb{R}
^{n}$\ of any Lipschitz continous function $f:%
\mathbb{R}
\rightarrow 
\mathbb{R}
^{n}$ with Lipschitz constant $c>0$, \textit{i.e.} satisfying $\left\Vert
f(t)-f(t^{\prime })\right\Vert \leq c\left\vert t-t^{\prime }\right\vert $
for all $t,t^{\prime }\in 
\mathbb{R}
$. In other words, a world line is a set of points $C\subseteq 
\mathbb{R}
^{n}\times 
\mathbb{R}
$ such that for every $t\in $\ $%
\mathbb{R}
$ there is one and only one $\mathbf{x}\in 
\mathbb{R}
^{n}$ with $\mathbf{x}t=(\mathbf{x},t)\in C$ and where $f:t\mapsto \mathbf{x}
$ is Lipschitz continuous with constant $c.$\ These serve to describe the
evolution in physical space-time of particles with or without mass.

\bigskip

\textbf{Proposition 2.1 \ }\textit{World lines are precisely the maximal
chains in the causality order of space-time.}

\bigskip

\textbf{Proof.} Every world line\ is a chain due to the Lipschitz condition,
and it is a maximal chain, as the addition of any other point would result
in two points with the same time component, which would be uncomparable in
the causality order.

Conversely, let $C$ be a maximal chain. Observe first that $C$ must be
topologically closed, and that no two points in $C$ can have the same time
component. Let $T$ be the set of time components of the points in $C.$ The
map $f$\ associating to each $t\in T$ \ the unique $\mathbf{x\in }$ $%
\mathbb{R}
$ such that $\mathbf{x}t\in C$ is Lipschitz\ continuous with constant $c$,
consequently $T$ is also closed in $%
\mathbb{R}
.$\ Let us show that $T=%
\mathbb{R}
$. If $T$ had a largest element $s,$ then adding the point $(f(s),s+1)$ to $%
C $ would result in a larger chain, which is impossible. Therefore $T$ is
not bounded above, and similarly, $T$ is not bounded below. If some real
number $r$ failed to belong to $T$, then $T$ would have a greatest element $%
a $\ smaller than $r,$ and a smallest element $b$ greater than $r.$ Adding
to $C$ any internal point of the line segment between $((f(a),a)$ and $%
((f(b),b)$\ would result in a larger chain, which is again impossible. \ \ \ 
$\square $

\bigskip

The following shows that antichain cutsets are in some sense "space-like
hypersurfaces", including all space-like hyperplanes but not requiring
linearity or smoothness. (A hyperplane is \textit{space-like} if it is a
causality antichain.)

\bigskip

\textbf{Proposition 2.2 \ }\textit{In the causality order} $\leq _{c}$ 
\textit{of }$(n+1)$-\textit{dimensional space-time} $%
\mathbb{R}
^{n+1}=%
\mathbb{R}
^{n}\times 
\mathbb{R}
,$ \textit{if} \textit{a set} $A$ \textit{of points constitutes an antichain
cutset, then it is the graph of a} \textit{Lipschitz-continuous function} $h:%
\mathbb{R}
^{n}\longrightarrow 
\mathbb{R}
$ \textit{with strict constant} $c^{-1}$ (i.e. $\left\vert h(\mathbf{x})-h(%
\mathbf{x}^{\prime })\right\vert <c^{-1}\left\Vert \mathbf{x}-\mathbf{x}%
^{\prime }\right\Vert $ \textit{for all }$\mathbf{x},\mathbf{x}^{\prime }\in 
\mathbb{R}
^{n}$).

\bigskip

\textbf{Proof.} \ If $A$ is an antichain cutset, then it intersects at
exactly one point each world line with fixed space component, and therefore
it is the graph of a map $%
\mathbb{R}
^{n}\longrightarrow 
\mathbb{R}
.$ As any two points on this graph are unrelated by causality, the Lipschitz
condition must hold.\ \ $\square $

\bigskip

\textit{Remark.} The converse does not hold, although the graph $A$ of a
Lipschitz continuous function is an antichain.

\bigskip

\textbf{Proposition 2.3 } \textit{The causality order} $\leq _{c}$ \textit{of%
} $(n+1)$-\textit{dimensional space-time }$%
\mathbb{R}
^{n+1}=%
\mathbb{R}
^{n}\times 
\mathbb{R}
$ \textit{is a gradable partial order, in which a set of points} $A\subseteq 
\mathbb{R}
^{n+1}$ \textit{is an antichain} \textit{cutset if and only if it is a level
set with respect to some grading.}

\bigskip

\textbf{Proof.} \ Gradability of the causality order is obvious, e.g. by $%
\mathbf{x}t\mapsto t.$ In fact, different gradings can be based on different
antichain cutsets, constructed as follows. For any antichain cutset $A$
there is a unique map $g:%
\mathbb{R}
^{n+1}\longrightarrow 
\mathbb{R}
$ such that for each $\mathbf{x}t\in $ $%
\mathbb{R}
^{n+1}$, the point $(\mathbf{x},t-g(\mathbf{x}t))$ belongs to $A$. This map
is surjective onto $%
\mathbb{R}
$ and it is a grading of the causality partial order for which $A$ is the
level zero. Thus every antichain cutset is a level set of some grading, and
the converse is true in all graded posets. \ $\square $

\bigskip

\bigskip

\textbf{3 \ Partial orders invariant under space isometries and space-time
dilations}

\bigskip

For any given partial order $\leq $ on any set, another, weaker order $\leq
^{\prime }$ on the same set is defined by 
\begin{equation*}
a\leq ^{\prime }b\ \ \ \ \ \Leftrightarrow \ \ \ \ \ a\leq b,\text{ and the
interval }\left[ a,b\right] \text{ is not a chain unless }a=b
\end{equation*}%
Generally the original order cannot be reconstructed from this weaker oder.
However, as apparent in Zeeman $\left[ Z\right] $, for all $n\geq 1$ the
causality order $\leq _{c}$ can be reconstructed from the \textit{subluminal
causality} order $\leq _{c}^{\prime }$ and they have the same automorphisms.
We have 
\begin{equation*}
\mathbf{u}\leq _{c}\mathbf{v}\text{ \ \ \ \ }\Leftrightarrow \text{ \ \ \ \ }%
\mathbf{u}\leq _{c}^{\prime }\mathbf{v}\text{ or }(\forall \mathbf{w\neq u,v}%
\ \ \ \mathbf{v}\leq _{c}^{\prime }\mathbf{w\Rightarrow u}\leq _{c}^{\prime }%
\mathbf{w)}
\end{equation*}

\bigskip

Let $\mathbf{x}t\in 
\mathbb{R}
^{n}\times 
\mathbb{R}
$ be any forward light-like vector, \textit{i.e.} not $\mathbf{0}0$ and such
that $\left\Vert \mathbf{x}\right\Vert =t>0$. Then the set $\left\{ \mathbf{0%
}r:r<0\right\} \cup \left\{ \mathbf{x}s:t\leq s\right\} $ is a maximal chain
in the order of subluminal causality, and it avoids any antichain of
subluminal causality that contains $\mathbf{0}0.$ Applying translation to
the origin, this shows that, in the order of subluminal causality, for any
antichain there are maximal chains that avoid it. Consequently we have:

\bigskip

\textbf{Proposition 3.1} \ \textit{In the order of subluminal causality
there are no antichain cutsets, and the subluminal causality poset cannot be
graded. \ }$\square $

\bigskip

For every positive real constant $c$, both causality $\leq _{c}$ and
subluminal causality $\leq _{c}^{\prime }$ are invariant under \textit{space
isometries} (i.e. under transformations of $%
\mathbb{R}
^{n+1}=%
\mathbb{R}
^{n}\times 
\mathbb{R}
$ of the form $\mathbf{x}t\mapsto \mathbf{y}t$ where $\mathbf{x\mapsto y}$
is an isometry of Euclidean $n-$space), and also invariant under all
space-time dilations \ $\mathbf{x}t\mapsto (r\mathbf{x},rt)$, $r>0.$
Conversely, let $\leq $ be any partial ordering of $%
\mathbb{R}
^{n+1}=%
\mathbb{R}
^{n}\times 
\mathbb{R}
$ that is invariant under space isometries and space-time dilations. The set
of elements greater or equal to $\mathbf{0}0$ is then a cone $C$ invariant
under \textit{space rotations} (space isometries fixing $\mathbf{0}0$), and
this cone determines the order $\leq $ by the condition $\mathbf{u\leq v}$ $%
\Leftrightarrow $ $\mathbf{v}-\mathbf{u\in }C.$ If $C$ is topologically
closed, then it is the forward causality cone of $\leq _{c}$ for some
positive constant $c$ or its negative, the backward causality cone. If $C$
is not closed, then there can be two cases. In the first case $C$ is the
forward subluminal causality cone of $\leq _{c}^{\prime }$ for some positive
constant $c$ or its negative, the backward subluminal causality cone. In the
second case $C$ is $\left\{ \mathbf{0}t:t>0\right\} \cup \left\{ \mathbf{0}%
0\right\} $ or its negative $\left\{ \mathbf{0}t:t<0\right\} \cup \left\{ 
\mathbf{0}0\right\} $: the corresponding partial orders are the forward and
backward \textit{temporal orderings} of Galilean spacetime, and they have
too many automorphisms to which no physical meaning is attributed. Clearly
we have\textit{\ }$\mathbf{u\leq v}$ under temporal ordering if and only if $%
\mathbf{u\leq }_{c}\mathbf{v}$ under the causality ordering for all positive 
$c$, or equivalently, if and only if $\mathbf{u\leq }_{c}^{\prime }\mathbf{v}
$ under subluminal causality for all positive $c$. In that sense temporal
ordering is "subluminal causality at infinite speed of light".

\bigskip

\textbf{Proposition 3.2} \ \textit{The only partial orderings of} $(n+1)-$%
\textit{dimensional spacetime that are} \textit{invariant under space
isometries and space-time dilations are causality and subluminal causality
with various light speed parameters }$c$\textit{, the temporal ordering, and
the reverse orders of these.} \textit{\ }$\square $

\bigskip

\textit{Remark and some recall of definitions. }We now assume that $c=1.$
The causality and subluminal causality order, but not the temporal order,
are also invariant under hyperbolic rotations, called \textit{boosts}
(conjugates of the form $rbr^{-1}$, where $r$ is a space-time isometry
fixing the origin, and $b$\ is a\ linear space-time transformation with two
reciprocal positive eigenvalues corresponding to the eigenvectors $%
(1,0,...,0,1)$ , $(-1,0,...,0,1)$ and fixing all the standard unit vectors
other than $(1,0,...,0,0)$ and $(0,0,...,0,1).$ Boosts composed with space
isometries preserving the origin (i.e. with \textit{space rotations}) make
up \textit{Lorentz transformations}, which constitute the \textit{Lorentz
group} of linear transformations preserving the \textit{Minkowski norm} $%
x_{1}^{2}+...+x_{n}^{2}-t^{2}$ of space-time. Lorentz transformations
composed with \textit{translations} of space-time ($\mathbf{x}t$ $\mapsto 
\mathbf{x}t+\mathbf{d}s$, for some fixed $\mathbf{d}s\in 
\mathbb{R}
^{n+1}$) make up the \textit{Poincar\'{e} group} of transformations and
further enlargement with space-time dilations generates the group of all
automorphisms of causality order (or subluminal causality order). This fact
amounts to the Alexandrov-Zeeman Theorem, for which a variant proof based on
order-theoretical notions is proposed in the next section.

\bigskip

\bigskip

\textbf{4 Variant proof of the Alexandrov-Zeeman Theorem}

\bigskip

Throughout this section we continue to assume that the light speed constant $%
c$\ is $1$ (the choice of kilometres and seconds as measuring units being
arbitrary).

The result -- an exact statement of which is given below - was first
obtained by Alexandrov ($\left[ A1,AO\right] ,$ see also the later article $%
\left[ A-CJM\right] $), then proved independently by Zeeman $\left[ Z\right] 
$. \textit{Poincar\'{e} transformations} are Lorentz transformations
composed with space-time translations. We refer to the group generated by
Poincar\'{e} transformations and space-time dilations as the \textit{dilated
Poincar\'{e} group}. It is well known, and can be shown by simple linear
algebraic methods without reference to the order-theoretic properties of the
causality relation, that the dilated Poincar\'{e} group acts transitively on
each of the following sets:

\textit{(i)} the set of all space-time points,

\textit{(ii)} the set of \textit{light-like vectors}, the set of \textit{%
time-like} \textit{vectors}, and the set of \textit{space-like vectors}
(these being the set of non-null vectors with null, negative and positive
Minkowski norm, respectively),

\textit{(iii)} the set of \textit{optical lines}, the set of \textit{%
separation lines}, and the set of \textit{inertia lines} (being the
translates of 1-dimensional subspaces generated by light-like, time-like and
space-like vectors, respectively).

If a Lorentz transformation fixes the time-like standard unit vector $%
(0,\ldots ,0,1)$, then it must be a space isometry (indeed a space rotation).

Two further standard notions needed in the proof is that of \textit{optical
hyperplane} (hyperplane containing optical and separation lines, but no
inertia lines, defineable alternatively as the tangent hyperplanes of light
cones), and that of inertia plane (plane spanned by two intersecting optical
lines, these in fact always contain separation and inertia lines as well).
For references to some of these standard facts and terminology and a
physical perspective see e.g. Latzer $\left[ L\right] $, Moretti $\left[ M%
\right] $ or Urbantke $\left[ U\right] $. \newpage 

\textbf{Statement of the Alexandrov-Zeeman Theorem} $\left[ A1,A2,Z\right] $

\textit{In }$n+1$ \textit{dimensional spacetime, the group of automorphisms
of the causality order} $\leq _{c}$ \textit{is the group generated by the
Poincar\'{e} group and the dilations of spacetime.}

\bigskip

\textbf{Proof.} It is obvious that all Poincar\'{e} transformations
(generated by boosts, space isometries and spacetime translations), as well
as all dilations of spacetime, preserve causality.

Also it is clear that the order-automorphisms of causality and subluminal
causality are the same, each of these order relations being definable in
terms of the other.

The main task of the proof, as in the proofs of Alexandrov $\left[ A-CJM%
\right] $ and Zeeman $\left[ Z\right] $, is to establish linearity as a
consequence of order-preservation. However, continuity is not part of the
argument proposed here, as opposed to $\left[ A-CJM\right] $, and linearity
on optical lines does not play ther extensive preliminary role in the proof
that it does in $\left[ Z\right] $. Instead the following Observations
provide characterizations of all the three types of lines in spacetime
(optical, separation and inertia lines). Observations 3 and 5 express simple
properties of the linear algebraic structure of spacetime, reducing the
notions of separation line and inertia lines to that of order-theoretically
characterized optical hyperplanes and inertia planes.

\bigskip

\textit{Observation 1.} Optical lines are precisely the maximal order-convex
chains of the causality order.

\bigskip

\textit{Observation 2.} A set of points in spacetime is an optical
hyperplane if and only if it is the union of all members of some equivalence
class of optical lines, two optical lines being considered equivalent when
they have the same ,,subluminal causality neighborhood\textquotedblright\ in
the following graph theoretical sense. Referring to the set of points in a
graph (nodes, vertices) adjacent to at least one member of a set of points $%
S $ as the ,,neighborhood of $S$\textquotedblright , the subluminal
causality neighborhood of an optical line $L$ is the neighborhood of $L$ in
the comparability graph of the subluminal causality order. This neighborhood
is in fact always the set-theoretical complement in spacetime of the unique
optical hyperplane containing $L$.

\bigskip

\textit{Observation 3.} Separation lines are precisely the minimal non-empty
non-singleton intersections of optical hyperplanes.

\bigskip

\textit{Observation 4.} \ A set of points in space-time is an inertia plane
if and only if it is the union of two distinct intersecting optical lines $%
K,L$ and all separation lines meeting $K\cup L$ in precisely two points.

\bigskip

\textit{Observation 5.} \ A set of points $L$\ in space-time is an inertia
line if and only if it is the non-empty intersection of two distinct inertia
planes and $L$\ is neither an optical line nor a separation line.

\bigskip

From these Observations it is clear that all causal automorphisms map lines
to lines, i. e. they are affine transformations (linear transformations
composed with a translation). From here the proof can be concluded
essentially as in the last few lines of $\left[ Z\right] $.

To be explicit, take an arbitrary causal automorphism $k$. Then $k$ maps the
standard affine basis (the standard unit vectors $\mathbf{e}_{i}$ and the
null vector $\mathbf{0}$) to some affine basis $k(\mathbf{0}),k(\mathbf{e}%
_{1}),\ldots .,k(\mathbf{e}_{n}),k(\mathbf{e}_{n+1})$.

Let $d$ be the translation mapping $k(\mathbf{0})$ back to $\mathbf{0}$, so $%
dk(\mathbf{0})=\mathbf{0}$. Then $dk(\mathbf{e}_{1}),\ldots .,dk(\mathbf{e}%
_{n}),$ are linearly independent space-like vectors and $dk(\mathbf{e}%
_{n+1}) $ is a time-like vector independent of them. There is a boost $b$
and a space-time dilation $a$ such that $badk(\mathbf{e}_{n+1})=(\mathbf{e}%
_{n+1})$, the vectors $badk(\mathbf{e}_{1}),\ldots .,badk(\mathbf{e}_{n})$
are linearly independent space-like vectors, and $\mathbf{e}_{n+1}$ is
independent of them. At this point we can conclude, as in $\left[ Z\right] $%
, that $badk$ is a Lorentz transformation. As it leaves $\mathbf{e}_{n+1}$
fixed, it must be a space isometry. Since this isometry as well as the
transformations $b,a,d$ belong to the dilated Poincar\'{e} group, so does $k$%
. \ \ \ $\square $

\bigskip

\bigskip

\textbf{Acknowledgements. }\ This work, undertaken while the author was at
the Tampere University of Technology in Finland, has been co-funded by Marie
Curie Actions (European Uniun) and supported by the National Development
Agency (NDA) of Hungary and the Hungarian Scientific Research Fund (OTKA),
within a project hosted by the University of Miskolc, Department of
Analysis. The work was also completed as part of the TAMOP-4.2.1.B.-
10/2/KONV-2010-0001 project at the University of Miskolc, with support from
the European Eunion, co-financed by the European Social Fund.

The author wishes to thank S\'{a}ndor Radeleczki and Mikl\'{o}s Ront\'{o}
for useful comments and discussions. \newpage 

\textbf{References}

\bigskip

[A1] A.D. Alexandrov, On Lorentz transformations, Sessions Math. Seminar,
Leningrad Section of the Mathematical Institute, 15 September 1949
(abstract, in Russian)

\bigskip

[AO] A.D. Alexandrov, V.V. Ovchinnikova, Note on the foundations of
relativity theory, Vestnik Leningrad Univ. 11 (1953) 95-100 (in Russian)

\bigskip

[A-CJM] A.D. Alexandrov, A contribution to chronogeometry, Canadian J. Math.
19 (1967) \ 1119-1128

\bigskip

[FW] S. Foldes, R. Woodroofe, Antichain cutsets of strongly connected
posets, Order 30 (2) 351-361 (2013)

\bigskip

[L] R.W. Latzer, Non-directed light signals and the structure of time, 
\textit{in} Space, Time and \ Geometry, P. Suppes (ed.) D. Reidel 1973, pp.
321-365

\bigskip

[M] V. Moretti, The interplay of the polar decomposition theorem and the
Lorentz group, in Lecture Notes of Seminario Interdisciplinare di Matematica
5-153 (2006) 18 pages, also arXiv:math-ph/0211047v1 at www.arxiv.org

\bigskip

[FW] I. Rival, N. Zaguia, Antichain cutsets, Order 1 (3) 235-247 (1985)

\bigskip

[U] H.K. Urbantke, Lorentz transformations from reflections: some
applications, Found. Phys. Lett. 16 (2003) 111-117

\bigskip

[Z] E.C. Zeeman, Causality implies the Lorentz group, J. Mathematical
Physics 5 (1964) \ 490-493

\bigskip

\bigskip

\bigskip

\end{document}